\documentclass[aps,longbibliography,superscriptaddress,twocolumn,10pt]{revtex4-1}

\usepackage{amsmath}
\usepackage{amssymb}
\usepackage{color}

\usepackage{graphicx}

\usepackage{subfigure}

\DeclareMathAlphabet{\mathitbf}{OML}{cmm}{b}{it}

\begin{document}
\title{Sprays from droplets impacting a mesh}

\author{Stefan Kooij}
\affiliation{Van der Waals-Zeeman Institute, University of Amsterdam, Science Park 904, Amsterdam, Netherlands}
\author{Ali Mazloomi Moqaddam}
\affiliation{Chair of Building Physics, Department of Mechanical and Process Engineering, ETH Zurich, 8092 Zurich, Switzerland}
\affiliation{Laboratory for Multiscale Studies in Building Physics, Empa, Swiss Federal Laboratories for Materials Science and Technology, 8600 D\"{u}bendorf, Switzerland}
\author{Thijs C. de Goede}
\affiliation{Van der Waals-Zeeman Institute, University of Amsterdam, Science Park 904, Amsterdam, Netherlands}
\author{Dominique Derome}
\affiliation{Laboratory for Multiscale Studies in Building Physics, Empa, Swiss Federal Laboratories for Materials Science and Technology, 8600 D\"{u}bendorf, Switzerland}
\author{Jan Carmeliet}
\affiliation{Chair of Building Physics, Department of Mechanical and Process Engineering, ETH Zurich, 8092 Zurich, Switzerland}
\author{Noushine Shahidzadeh}
\affiliation{Van der Waals-Zeeman Institute, University of Amsterdam, Science Park 904, Amsterdam, Netherlands}
\author{Daniel Bonn}
\affiliation{Van der Waals-Zeeman Institute, University of Amsterdam, Science Park 904, Amsterdam, Netherlands}

\begin{abstract}
\noindent
In liquid spray applications, the sprays are often created by the formation and destabilization of a liquid sheet or jet. The disadvantage of such atomization processes is that the breakup is often highly irregular, causing a broad distribution of droplet sizes. As these sizes are controlled by the ligament corrugation and size, a monodisperse spray should consist of ligaments that are both smooth and of equal size. A straightforward way of creating smooth and equally sized ligaments is by droplet impact on a mesh. In this work we show that this approach does however not produce monodisperse droplets, but instead the droplet size distribution is very broad, with a large number of small satellite drops. We demonstrate that the fragmentation is controlled by a jet instability, where initial perturbations caused by the injection process result in long-wavelength disturbances that determine the final ligament breakup. During destabilization the crests of these disturbances are connected by thin ligaments which are the leading cause of the large number of small droplets. A secondary coalescence process, due to small relative velocities between droplets, partly masks this effect by reducing the amount of small droplets. Of the many parameters in this system, we describe the effect of varying the mesh size, mesh rigidity, impact velocity, wetting properties, keeping the liquid properties the same by focusing on water droplets only. We further perform Lattice Boltzmann modeling of the impact process that reproduces key features seen in the experimental data.  
\end{abstract}

\maketitle

\section{introduction}
For many applications the atomization or spraying of a liquid is of paramount importance: from drug administration, printing, spray drying, to agriculture and firefighting; in all cases the droplet sizes play an important role. Usually, the spray is formed by a nozzle, first forming a liquid sheet or jet, that subsequently destabilizes to break up in columnar liquid structures, called ligaments. These ligaments further destabilize through the Rayleigh-Plateau instability driven by the surface tension, to form the final droplets of the spray \citep{strutt1878}. For Newtonian fluids the destabilization and breakup of these ligaments is by now well understood \citep{villermaux2007fragmentation,fraser1963atomization,dombrowski1963aerodynamic,villermaux2002life}. The distribution of droplet sizes is set by the initial ligament size and the ligament corrugation, where less or more corrugated ligaments result in less or more spread in droplet sizes \citep{villermaux2004ligament}. It has been shown that these parameters, i.e. ligament sizes and ligament corrugation, completely determine the final droplet size distribution in sprays \citep{kooij2018determines,villermaux2011drop}. The generic observation that sprays result in a wide distribution of drop sizes can then be understood: it is due to the random nature of the destabilization process of the sheet. This makes that ligaments are very corrugated and also frequently vary a lot in size, making droplet size distributions relatively broad.

To make sprays with monodisperse droplets, which are needed for many practical applications, the spraying process should therefore produce very smooth ligaments of equal size. We investigate such a design, in which a liquid is forced through a mesh. This design relies on the resulting ligaments having a uniform size determined by the dimensions of the pores, and the ligaments being relatively smooth. Fragmentation of a droplet impacting a mesh is a problem that occurs naturally in many situations \citep{kumar2018effect,zhang2018droplet}. In recent studies, the droplet impact on meshes was investigated \citep{soto2018droplet,ryu2017water}, and showed that indeed the ligaments are very smooth. They however did not consider the breakup mechanisms that determine the median droplet size and shape of the droplet size distribution as was done for regular sprays \citep{kooij2018determines,villermaux2011drop}

In this work we study the breakup of such smooth ligaments created by the impact of a droplet on a mesh. We find that the breakup is controlled by a jet instability, where initial perturbations caused by the injection process grow exponentially and fully determine the final fragmentation of the ligament. The perturbations typically cause a long wave disturbance, where the crests of the disturbances are connected by thinner ligaments, that break up in satellite-like drops that are much smaller than the main droplets. This spraying method therefore does not produce the desired mono-disperse sprays as one naively would expect; instead, the droplet size distribution consists of two characteristic peaks, one for the satellite droplets and the other for the main drops.  
\begin{figure*}[hbt!]
  \centering
  \subfigure{\includegraphics[width = 0.7\textwidth]{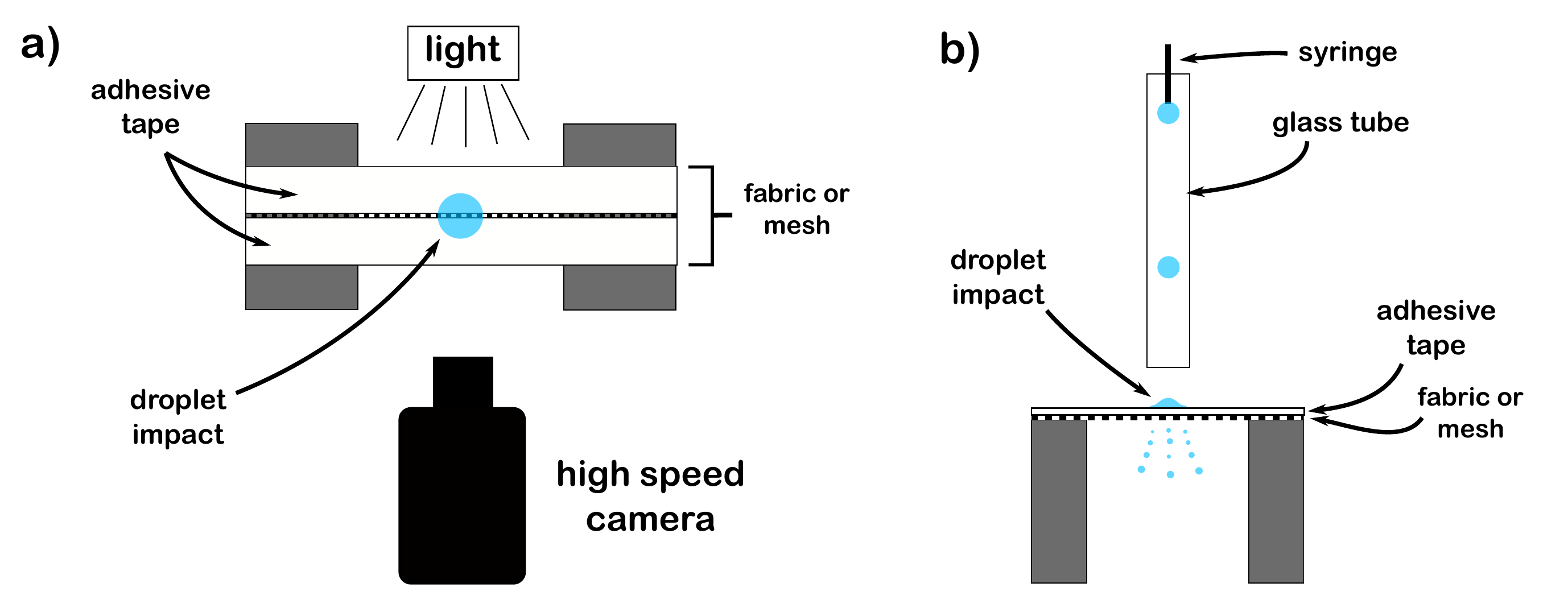}}
  \subfigure{\includegraphics[width = 0.28\textwidth]{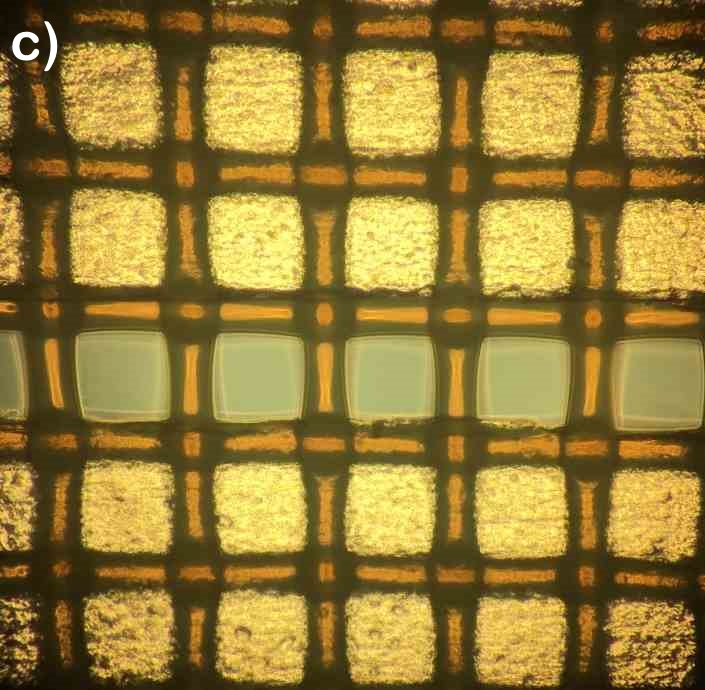}}
\caption{a) Schematic top (a) and side (b) views of the set-up (not to scale). A droplet (blue) impacts the middle of a fabric or mesh mounted onto two metal pillars. The tension in the mesh can be changed to alter the rigidity of the pores under study. Two pieces of adhesive tapes are applied to the mesh to leave only a single row of pores open. For experiments on a full mesh, these tapes are removed. Droplet impact is filmed with a high-speed camera (around 8000 fps) with back lighting. Droplets are produced by a blunt syringe needle, and travel through a glass tube, to ensure they fall on the exact same spot every time. c) Microscope picture of the fabric with one row of pores left open by taping the other holes closed. Diameter of the holes is 150 $\mu m$.\footnotesize }  
\label{fig:introduction_illustration}
\end{figure*}

Since the breakup dynamics are difficult to study due to the many ligaments that are created simultaneously during droplet impact, we also look at the fragmentation of a droplet falling on just a single row of pores (Fig.~\ref{fig:introduction_illustration}). Therefore, after the treatment of the experimental setup (Sec.~\ref{seq:experimental}), the result are divided into two parts: results for the impact on (regular) meshes (Sec.~\ref{seq:results_normal_meshes}) and results for single-row meshes (Sec.~\ref{sec:single_row_meshes}).

\section{Experimental}
\label{seq:experimental}

Two type of meshes where used: polyester fabrics of mesh size $45\;\mu m$, $106\; \mu m$, $150\; \mu m$ and a brass mesh of $300 \; \mu m$ and yarn diameters $40 \; \mu m$, $70\; \mu m$, $80\; \mu m$ and $150 \; \mu m$ respectively. The meshes of around 1 cm wide are spanned over a small gap (8 mm) between two metal pillars. In the case of the polyester fabric, the fabric is pulled tight across the gap to make it more rigid (Fig.~\ref{fig:introduction_illustration}). A high-speed camera films the breakup events in front of the gap with backlighting, with a frame rate of $\sim 8000$ fps (See Fig.~\ref{fig:overview_impacts} and Movies S2 and S3 in the Supplemental Materials). The height of the camera can be adapted, depending on which part of the dynamics needs to be captured. Droplets are created by a blunt-needle syringe and fall through a glass tube. This ensures that droplets are not affected by surrounding air currents and consistently impact the same spot. Excess water is removed with paper between droplet impact events, but the mesh is not completely dried. 

\begin{figure*}[htb!]
\centering
\includegraphics[width = 1.0\textwidth]{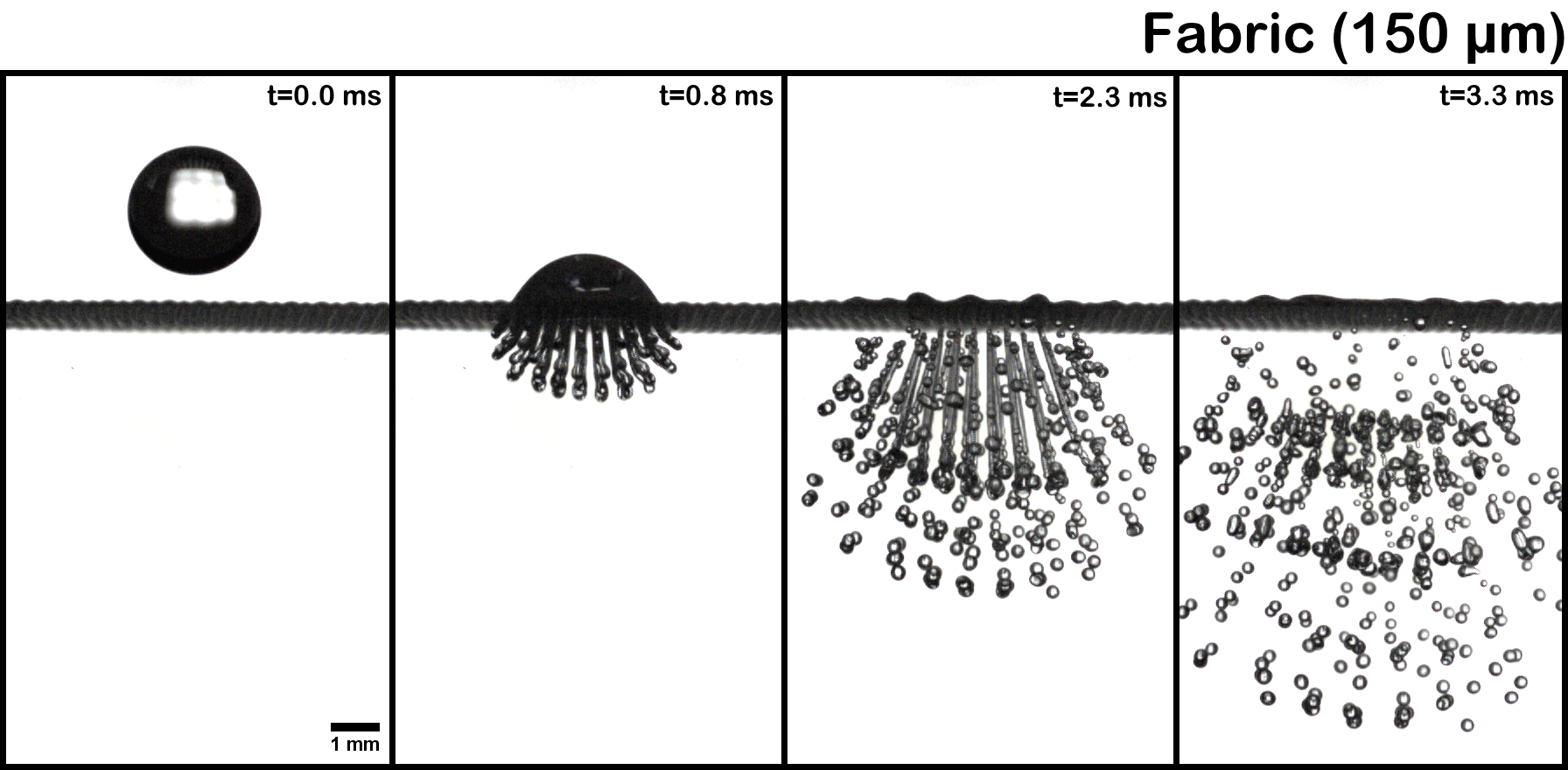}
\caption{Image sequence of a droplet impact on a mesh (150 $\mu m$ pore polyester fabric) with an impact velocity of 2.7 m/s (See also S1 in Supplemental Material). Due to the inertia of the droplet, liquid is being pushed through the mesh, creating many ligaments that break up to form the droplets of the spray. Due to overlapping trajectories, the breakup dynamics of the ligaments is difficult to analyze unlike the very similar case of a single row of pores (see Fig.~\ref{fig:overview_impacts})  \footnotesize }
\label{fig:whole_mesh_image_sequence}
\end{figure*}

When a droplet impacts a mesh, many ligaments and droplets are created at the same time (Fig.~\ref{fig:whole_mesh_image_sequence}), complicating studies of the dynamics. Therefore, we consider the impact of drops on a full mesh (Sec.~\ref{seq:results_normal_meshes}) as well as on a single row of pores  (Sec.~\ref{sec:single_row_meshes}) by covering most of the mesh using adhesive tape (see Fig.~\ref{fig:introduction_illustration}c). These so-called single-row meshes facilitate studying the dynamics in detail (Fig.~\ref{fig:overview_impacts}). We assume that the dynamics of the single-row meshes is similar to that of the case where no tape is applied, which appears to be correct when comparing high-speed camera footage of the spreading and ejection of droplets for both cases. 

The fragmentation of a droplet impacting a mesh involves a great number of system parameters, such as mesh size, initial drop size, surface tension, wetting properties of the liquid on the mesh material, yarn diameter, rigidity of the mesh, viscosity, impact velocity, etc. We keep the fluid properties the same by using only water, and vary many of the other system parameters at least to some extent. We note that even though the ligament shape and destabilization in most cases is strongly affected by varying these system parameters, the resulting droplet size distributions are similar. For the single-row experiments a drop height of $40\; cm$ was chosen, which is equivalent to an impact velocity of $2.7 \;m/s$. For the full-mesh experiments the drop height was varied, see Section \ref{sec:impact_velocity}.     

During fragmentation of a droplet, many smaller droplets are created that travel at different speeds and directions. This makes the measurement of drop sizes from a single picture problematic, since close to the mesh multiple droplets overlap each other, while further down, away from the mesh, not all droplets are inside one frame due to the different velocities with which they travel. To solve this issue, we constructed an algorithm that finds individual droplets from high-speed footage taken of the falling droplets. For the input of the program, the image sequences of the passing droplets are binarized and the position of each circular object in each frame is determined using ImageJ software. Because droplets (within the frame) travel at a constant velocity and have an almost straight trajectory, the position and size of individual droplets can easily be determined. In addition, shortly after fragmentation, many droplets coalesce due to droplets having relative velocities while moving along the same line. This is a complicating factor since the distribution changes over time, a phenomenon that is often ignored. We will show that this coalescence of droplets can significantly affect the size distribution (Sec.~\ref{sec:coalescence}). 

Unlike the case where a primary droplet impacts a full mesh, for the single-row meshes, droplet sizes can actually be measured by using just a single picture. This simplifies the analysis, but since each event contains a smaller number of droplets, around 100 droplet impacts are needed to obtain sufficient statistics.

\section{Results: drop impact on a mesh}
\label{seq:results_normal_meshes}
\begin{figure}[h!]
  \subfigure{\includegraphics[width = 0.38\textwidth]{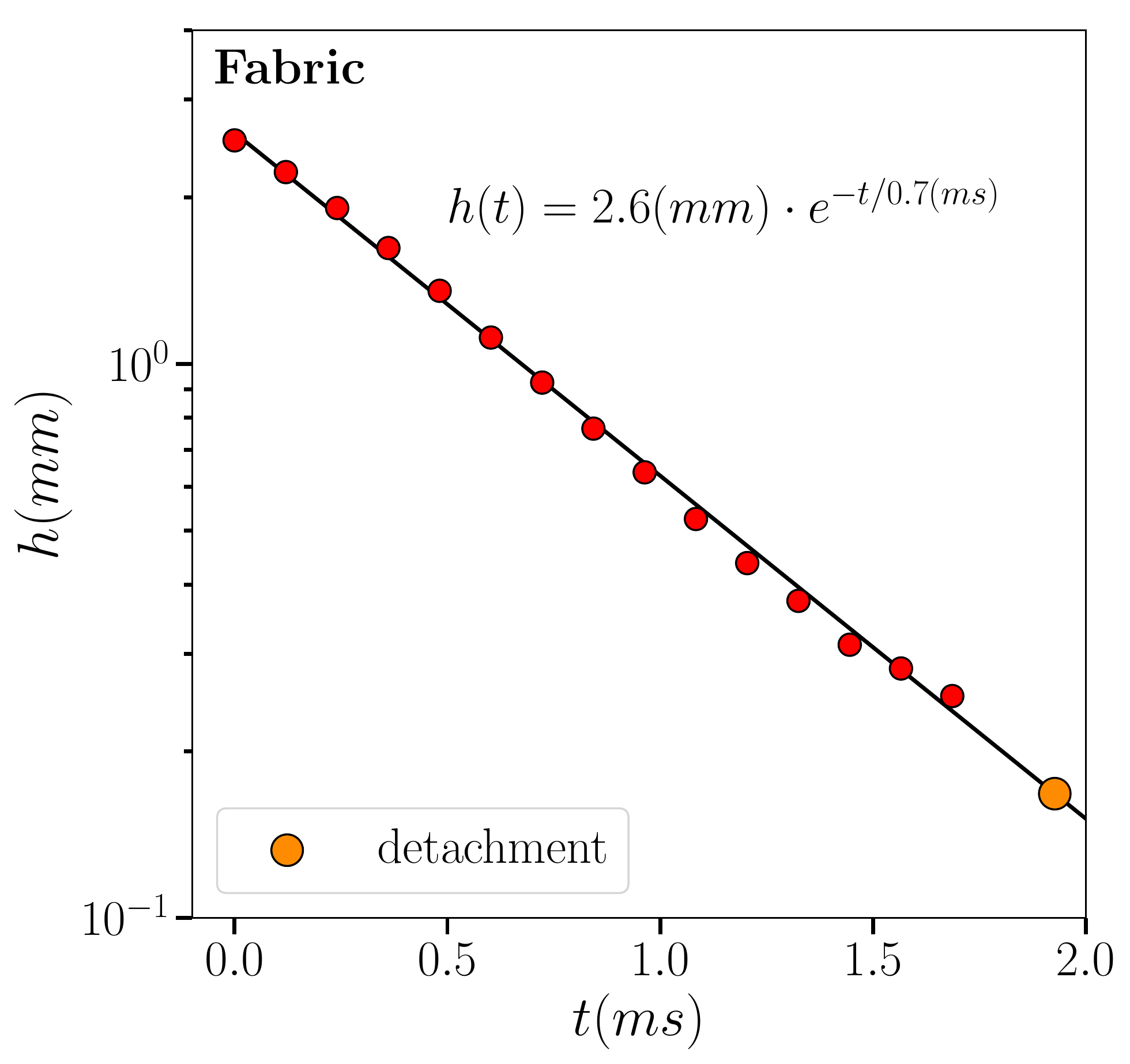}}
  \subfigure{\includegraphics[width = 0.08\textwidth]{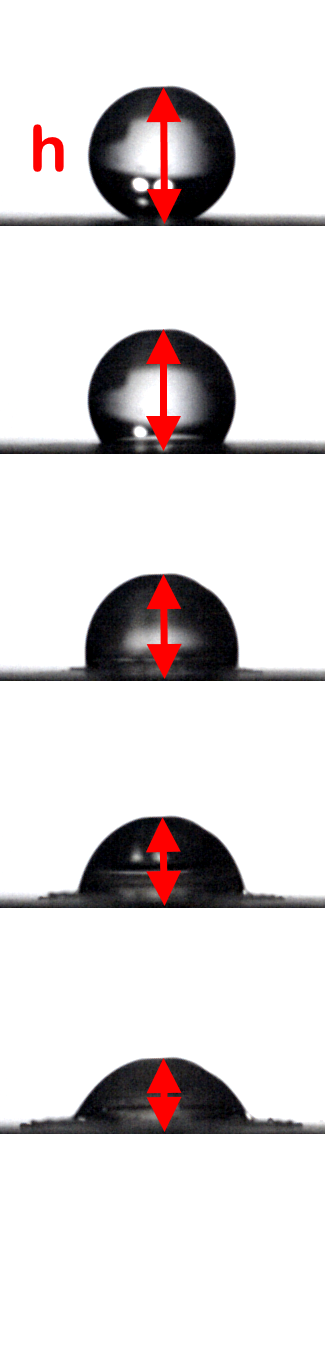}}
\caption{Height measurements of the hemi-spherical top of an impacting droplet of a single-row fabric (150 $\mu m$) with an impact velocity of 2.7 $m/s$. The data is fitted with an exponential which gives a characteristic time of 0.7 $ms$ that will be later used in the simulations. Very similar results were obtained for a full metallic mesh (300 $\mu m$). \footnotesize }  
\label{fig:exp_injection_speed}
\end{figure}

By measuring the height of the hemi-spherical part of the impacting droplet we find that, by approximation, the injection speed for the central jets slows down exponentially with time (Fig.~\ref{fig:exp_injection_speed}). 
This can be explained by equating the kinetic energy of the spreading droplet with the surface energy. Taking $D(t)$ to be the droplet diameter of the spreading droplet, $D_{0}$ the initial drop diameter, $\rho$ the fluid density and $\gamma$ the surface tension, one finds
\begin{eqnarray}
\rho D_{0}^{3} \left(\frac{d D}{d t}\right)^{2} \sim \gamma D^{2} \rightarrow \frac{d D}{D} \sim \sqrt{\frac{\gamma}{\rho D_{0}^{3}}} dt \rightarrow  \\ 
D \sim e^{\sqrt{\gamma/\rho D_{0}^{3}}}, \nonumber
\end{eqnarray}
and because of volume conservation we can write $h \sim D_{0}^{3}/D^{2}$, which gives  
\begin{equation}
h(t) \sim C e^{-2\sqrt{\gamma/\rho D_{0}^3}t}.
\end{equation}
This gives a characteristic timescale of approximately 0.5 $ms$, and since at $h(t\rightarrow 0) = C$ the prefactor should be of the order of the drop diameter $D_{0}$. Indeed these values agree well with the experimental fit parameters, even though this derivation ignores all the complex spreading dynamics. 

These results imply that the droplet formation mechanism can to a large degree be viewed as a simple system of a cylinder with piston, where the piston height decreases exponentially with time, and at the bottom of the cylinder there is a hole with a diameter equal to the pore size. To test this hypothesis, and to explore a situation with no vibrations (something that is unattainable experimentally), we performed Lattice Boltzmann simulations of such a system. As discussed in Sec.~\ref{sec:simulations}, we find that the simulated ligament formation, in which wave disturbances on the detached ligaments are absent, is very similar to our experimental results. 

From high-speed camera images we identify three stages in the fragmentation process that follows the impact. At first, the droplet impacts the mesh and the injection speed is constant. The destabilization of the resultant ligament is a pure jetting phenomenon, which results in the breakup of 1 to 3 droplets at the end of the jet. Secondly, the impacted droplet spreads on the mesh surface, slowing down the injection speed exponentially. Due to inertia, the ligaments starts to stretch and thin, until the ejection speed is so slow that the ligament detaches from the mesh. In the final stage, the remaining detached ligament destabilizes by the growth of initial perturbations. The wavelength of the resultant disturbance depends on the system parameters, but is otherwise completely deterministic.  

\subsection{Drop size distribution}
It is now well established that the breakup of ligaments of a Newtonian fluid is best described by a fragmentation-fusion scenario \citep{villermaux_coalescence_scenario}. The drop size distribution is given by a Gamma function
\begin{equation}
\Gamma(n,x)  = \frac{n^{n}}{\Gamma(n)}x^{n-1}e^{-nx},
\label{eq:gamma}
\end{equation}
where $x=d/\left<d\right>$,$d$ is droplet diameter, $\left<d\right>$ is the average droplet diameter, and $n$ is a parameter set by the ligament corrugation before destabilization. Very corrugated ligaments correspond to $n\approx\;$4-5 and result in a broad drop size distribution, while the most smooth ligaments would lead to $n=\infty$ leading to a delta peak. In more complicated spray formation processes, ligaments can also vary in size, in which case the drop size distribution is a compound gamma distribution \citep{villermaux2011drop,kooij2018determines}. In this case the size distribution of the ligaments themselves has to be taken into account as well, which further widens the drop size distribution.

It is clear from our experiments that the ligaments all have the same size, given by the mesh size. The ligament corrugation parameter $n$ can then easily be estimated from the high-speed camera footage. Considering the smoothness of the ligaments before breakup, $n$ should be very large, and the resulting distribution very narrow. However, this holds for ligaments with random initial perturbations, not for jets such as the ones created in our experiments. For such jets the breakup is largely deterministic, where the nature of the initial disturbance governs the final breakup.  

\begin{figure}[!h]
\centering
\includegraphics[width = 0.48\textwidth]{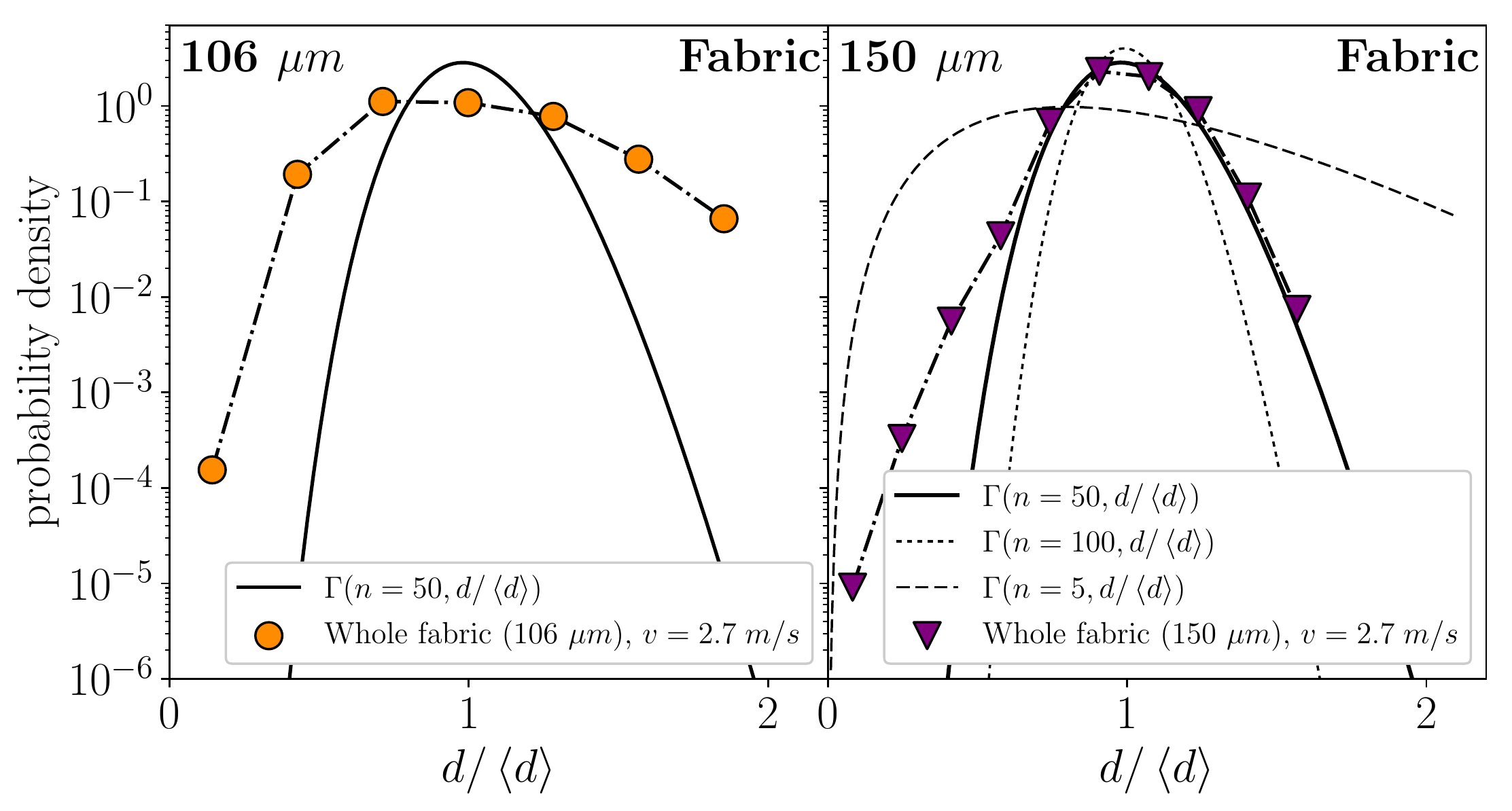}
\caption{Droplet size distributions measured for full meshes with pore diameters 106 $\mu m$ (left panel) and 150 $\mu m$ (right panel), and impact velocity $v= 2.7\;m/s$. Inserted are plots of $\Gamma(n=50,d/\left<d\right>)$, $\Gamma(n=100,d/\left<d\right>)$ and $\Gamma(n=5,d/\left<d\right>)$ showing that there is no reasonable fit for any parameter $n$. Still $n=50$ fits reasonably well except for small droplets. This value of $n$ is reasonable for such smooth ligaments as seen just before breakup.  \footnotesize }
\label{fig:distribution_whole_mesh_single_velocity}
\end{figure}

Using the tracking algorithm we measured the droplet sizes for the full-mesh case and polyester fabrics of 106 $\mu m$ and 150 $\mu m$ with an impact velocity of 2.7 $m/s$. In Figure \ref{fig:distribution_whole_mesh_single_velocity} shows the rescaled distributions. To compare the drop size distributions for ordinary ligaments, a plot of $\Gamma(n=50,d/\left<d\right>)$ is added to the graph together with plots for $n=5$ and $n=100$, which shows that the shape of the measured distribution does not fit well with a Gamma function. Still $n=50$ fits well, which is a reasonable estimate for smooth ligaments that vary a bit in thickness along their length such as in our experiments. Ligaments with corrugations $n > 100$ are already hard to distinguish from each other and can be considered to be ``straight ligaments". For both pore sizes, the shape of the distribution is considerably different from the fit line, especially for the 106 $\mu m$ fabric, showing that this fragmentation method actually performs rather poor compared to other atomization methods in terms of monodispersity of the drops. For the 106 $\mu m$ fabric, there is also an excess of large droplets compared to the 150 $\mu m$ fabric. This is likely due to the fact that for the 106 $\mu m$ fabric the polyester yarn diameter is smaller, making it more likely that some jets merge, creating bigger droplets.

The origin of the excess of small droplet becomes apparent when we look at the breakup of a droplet impacting a single row of pores, making the breakup mechanism visible. This will be further discussed in the second part of the results about single row meshes (Sec.~\ref{sec:single_row_meshes}).

\subsection{Some system variables}
Of the many parameters, we investigated several key ones, often only reporting the qualitative response for a specific set-up.  

\subsubsection{Impact velocity}
\label{sec:impact_velocity}
Droplet impact velocities were varied from 2.1 $m/s$ to 3.0 $m/s$ by changing the drop height for the case of a full mesh of polyester fabric. We did not consider lower speeds as they, depending on the other system parameters, do not always result in the droplet fragmenting, or result in in an insufficient number of fragments to compose a proper size distribution. For similar reasons, higher velocities are also not suitable, since then too many droplets are created, making image analysis problematic.  

\begin{figure}[!h]
\centering
\includegraphics[width = 0.48\textwidth]{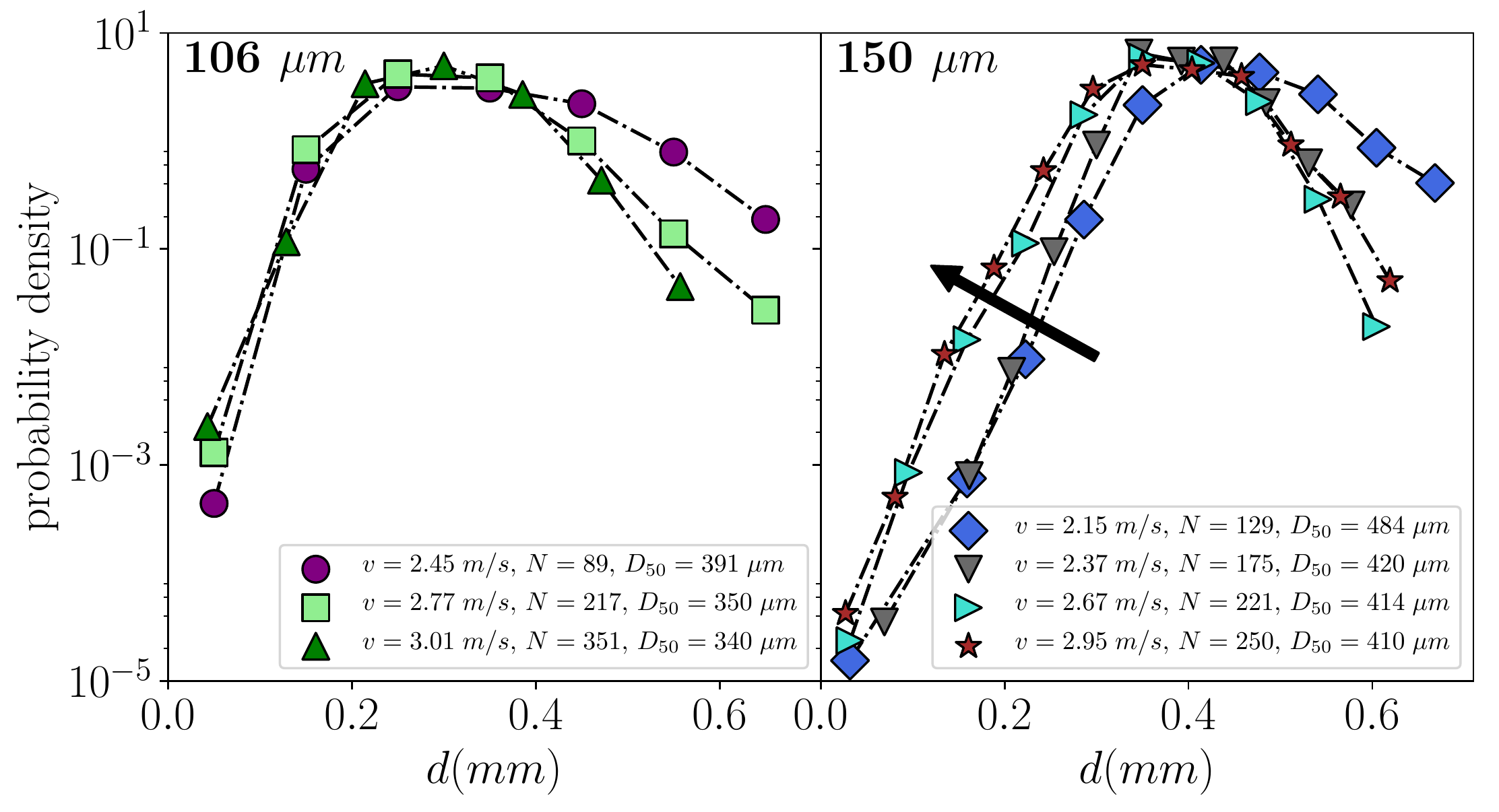}
\caption{Droplet size distributions for the impact on a full mesh of polyester fabric with pore size $106\;\mu m$ and $150\;\mu m$ for different impact velocities. Especially for the $150\;\mu m$ pore size, the average droplet size shifts to smaller droplets when the impact velocity increases as indicated by the arrow. The average amount of droplets created per event strongly increases with velocities.   \footnotesize }
\label{fig:distribution_whole_velocities}
\end{figure}

The results are shown in Fig.~\ref{fig:distribution_whole_velocities} and reveal that for increasing impact velocity the average amount of droplets per event rises and the size distribution shifts slightly to smaller droplets. 
This can be explained by the larger stretch ligaments experience at higher velocities, making them thinner on average, thereby reducing the average droplet size. The strong increase in the number of droplets with the increase in kinetic energy is mostly due to the increased mass transfer through the fabric, and not by the relatively small decrease in drop size. Besides the small shift there is also a reduction in the number of the largest droplets with the increase in impact velocity. These larger droplets are much larger than the pore size, suggesting that they originate from some type of merging, such as the coalescence of droplets or ligaments. Indeed, high-speed footage shows that ligaments sometimes can coalesce, especially when the yarn diameter is small such as for the fabric with $106\;\mu m$ pores. This merging is reduced for higher impact velocities, therefore decreasing the amount of the largest droplets. 

\label{sec:impact_velocity}
\subsubsection{Mesh size and rigidity}
For the fabric meshes we qualitatively varied the tension of the fabric between the two pillars. We find there is an increase in the mass transfer through the fabric, as could be expected; less tension causes a dampening of the impacting droplet, with less converted kinetic energy. Furthermore, we find that the destabilization as well as the detachment of the ligaments alters with the change of the rigidity of the mesh. However, the general features of the breakup as described in Sec.~\ref{sec:single_row_meshes} still hold, unless the tension is so low that penetration through the fabric is mostly inhibited. 

By increasing the pore size, one increases the drop size, although not linearly. The mean drop size is mostly controlled by the mean ligament diameter. The ligament diameter is however not only a function of the pore size, but also the amount of stretching, which in turn depends on other parameters. The exact relation between pore size and mean drop size therefore requires a more systematic approach, which is beyond the scope of this paper.  

\subsubsection{Wetting properties}

To vary the wetting properties of the mesh we either used a plasma treatment of the polyester fabric to make it hydrophilic, or used a hydrophobic spray to make the fabric hydrophobic. We observe that for the plasma-treated fabric the droplet impact does not lead to the formation of small ligaments anymore, but instead the water moves around the fabric wires during impact and forms one lump of water underneath the impact zone, which can be expected for more hydrophilic meshes. Making the mesh hydrophobic results in more droplets created without a significant change in droplet sizes. This shows that by changing the wetting properties the spraying performance can be altered, but if a state of jet formation is reached, this has little or no effect on the breakup process. It can be expected that especially for small yarn diameters, the hydrophobicity plays an important role in preventing of jets merging, which could be a useful tool in the production of mono-dispersed sprays.

\section{Results: single row of pores}
\label{sec:single_row_meshes}
\begin{figure*}[htbp!]
\centering
\includegraphics[width = 1.0\textwidth]{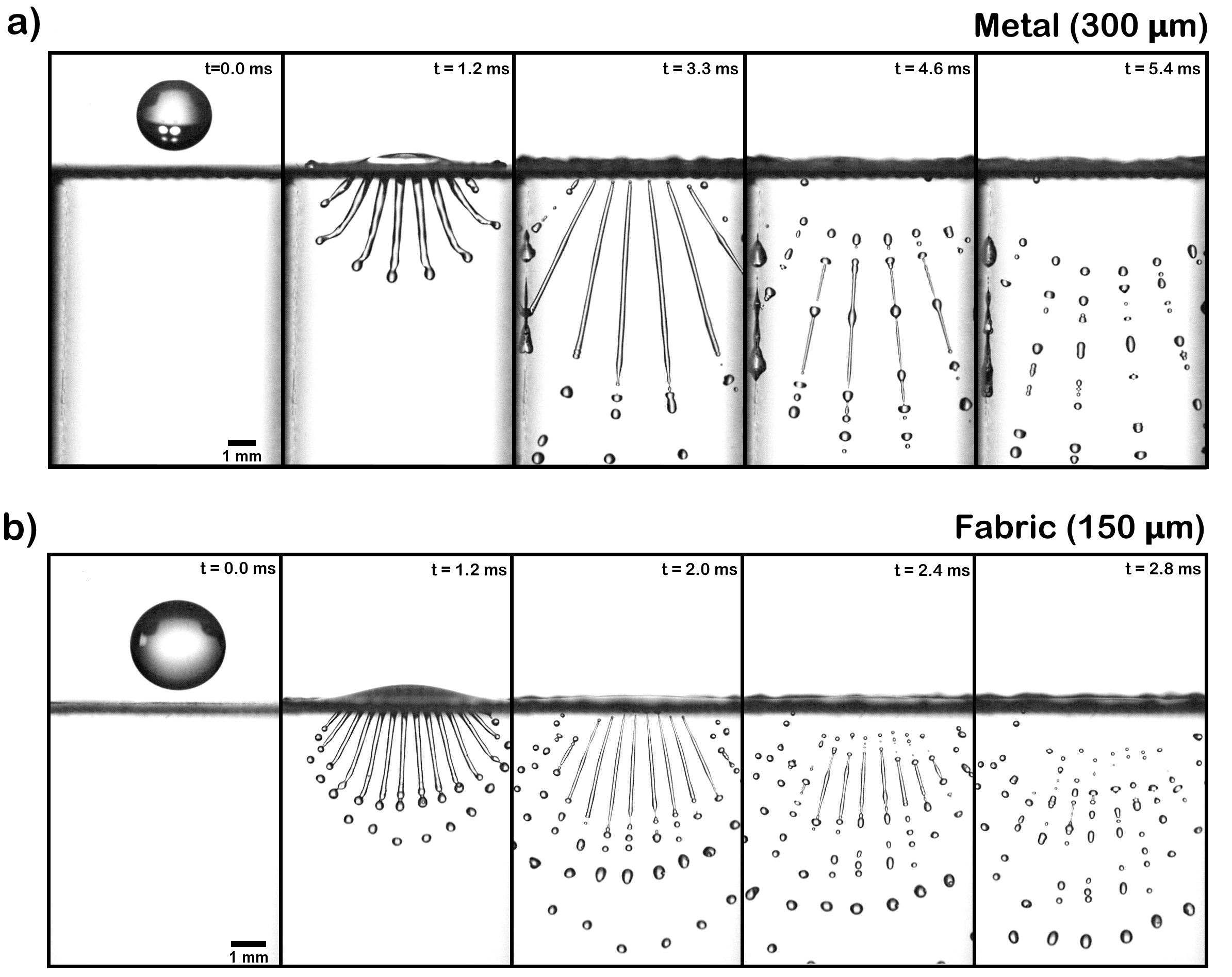}
\caption{Droplet impacts on meshes with a single row of open pores. Impact velocity is 2.7 m/s. a) Image sequence of a droplet impact on a single-row metallic (brass) mesh with pore size 300 $\mu m$ (see also Movie S2 in the Supplemental Materials). Due to the rigidity of the mesh it does not deform as a result of the droplet impact. b) Image sequence of a droplet impact on a single-row polyester fabric mesh with 150 $\mu m$ pores, that has previously been wetted, producing ligaments that are smooth and pointy at detachment (see also Movie S3 in the Supplemental Materials). \footnotesize }
\label{fig:overview_impacts}
\end{figure*}
Since the formation and destabilization of the ligaments created by the droplet impact on a normal mesh is hardly visible, the droplet impact on single-row meshes gives a number of  crucial insights. Although the presence of the tape undoubtedly has an effect on the flow of the impacting droplet, we expect that the general observations of the formation and fragmentation of the resultant jets are still applicable to the full-mesh case as visually the ligament formation and breakup is very similar between the two cases. Fig. \ref{fig:overview_impacts} shows the image sequence of the droplet impact for a 300 $\mu m$ metal mesh (a) and a 150 $\mu m$ polyester fabric (b).

\begin{figure}[!htb]
\centering
\includegraphics[width = 0.48\textwidth]{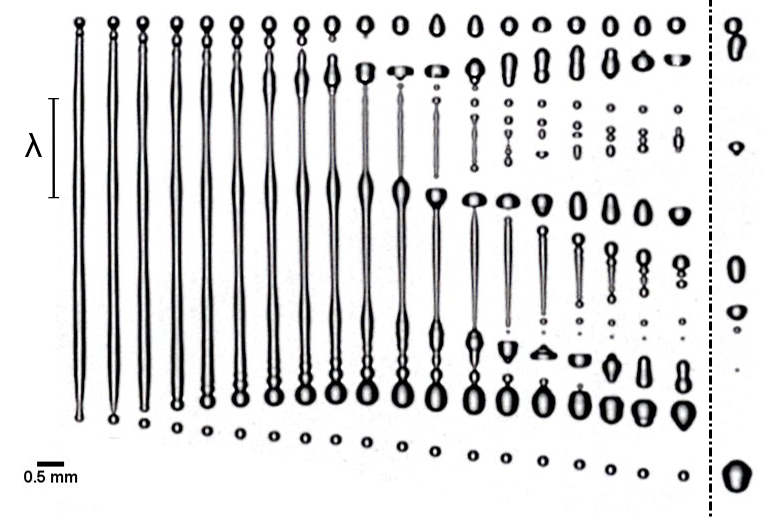}
\caption{Sequence of snapshots of a ligament breaking up aligned by the topmost droplet. The ligament was created by a droplet impact on a single-row metallic mesh. A clear wave disturbance can be observed, with a wavelength $\lambda$ of 1.8 $mm$. The instability grows with time, with the crests of the waves being connected by thinning ligaments that detach and form satellite droplets. The time between frames is 120.46 $\mu s$, giving a total time of 2.17 $m s$ between the first and second last frame. The last frame is taken at a later time of 4.22 $ms$, showing the effect of coalescence; multiple droplets have fused and the above two droplets are going to coalesce. It is clear that the stretching stops after detachment from the mesh. In fact, the ligament contracts a little before destabilizing.  \footnotesize }
\label{fig:breakup_sequence_metallic_mesh}
\end{figure}

\begin{figure}[!htb]
\centering
\includegraphics[width = 0.48\textwidth]{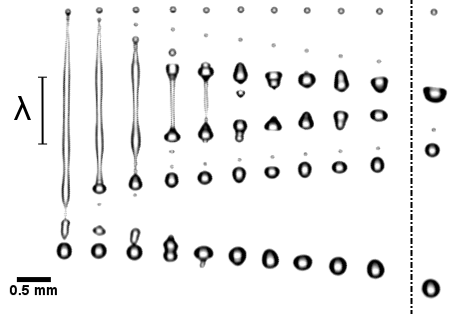}
\caption{Sequence of snapshots of a ligament breaking up aligned by the topmost droplet. The ligament was created by a droplet impact on a single-row polyester fabric. The wave disturbance can clearly be seen, with a wavelength $\lambda$ of 1 $mm$. The time between frames is 135.13 $\mu s$, giving a total time of 1.22 $ms$ between the first and second last frame. The last frame is taken at a later time of 2.03 $ms$, showing the effect of coalescence.   \footnotesize }
\label{fig:breakup_sequence}
\end{figure}

From this Figure we find that the droplet fragmentation can be divided in three stages. First, the droplet impact results in liquid being injected through the mesh at a relatively constant speed. The thus formed jet destabilizes at the tip, forming about three droplets that have a size of the order of the pore size. Next, the droplet starts to spread and the injection speed slows down quickly. Due to inertia, the slowing injection process causes the ligament to stretch and thin until the injection speed is so low that it detaches from the mesh. We find that the wetting properties have a significant impact on the detachment. If e.g. the polyester fabric is dried with hot air between droplet impact events, the detachment from the mesh is impeded, creating large droplets at the detachment point. Finally, the detached ligament destabilizes and breaks up in droplets. Figures \ref{fig:breakup_sequence_metallic_mesh} and \ref{fig:breakup_sequence} show a sequential breakup of these ligaments for the metal and fabric mesh, respectively. The frames are equally spaced in time except for the right most frame which is the last frame recorded. These last frames show a clear secondary process of coalescence, as further discussed in Sec.~\ref{sec:coalescence}.

In the breakup sequences (Figs~\ref{fig:breakup_sequence_metallic_mesh} and \ref{fig:breakup_sequence}) as well as in Fig. \ref{fig:overview_impacts}, clear long-wavelength disturbances can be observed, at the same relative locations for the different jets and between droplet impact events, i.e. the breakup always happens at the same locations. When the disturbances start to grow, the crests swell, being connected by thin ligaments that eventually break up into smaller satellite-like drops. The initial waves on the surface of the ligaments therefore completely determine the breakup of the produced ligaments, causing an abundance of small droplets. 


\begin{figure}[htbp!]
  \subfigure{\includegraphics[width = 0.48\textwidth]{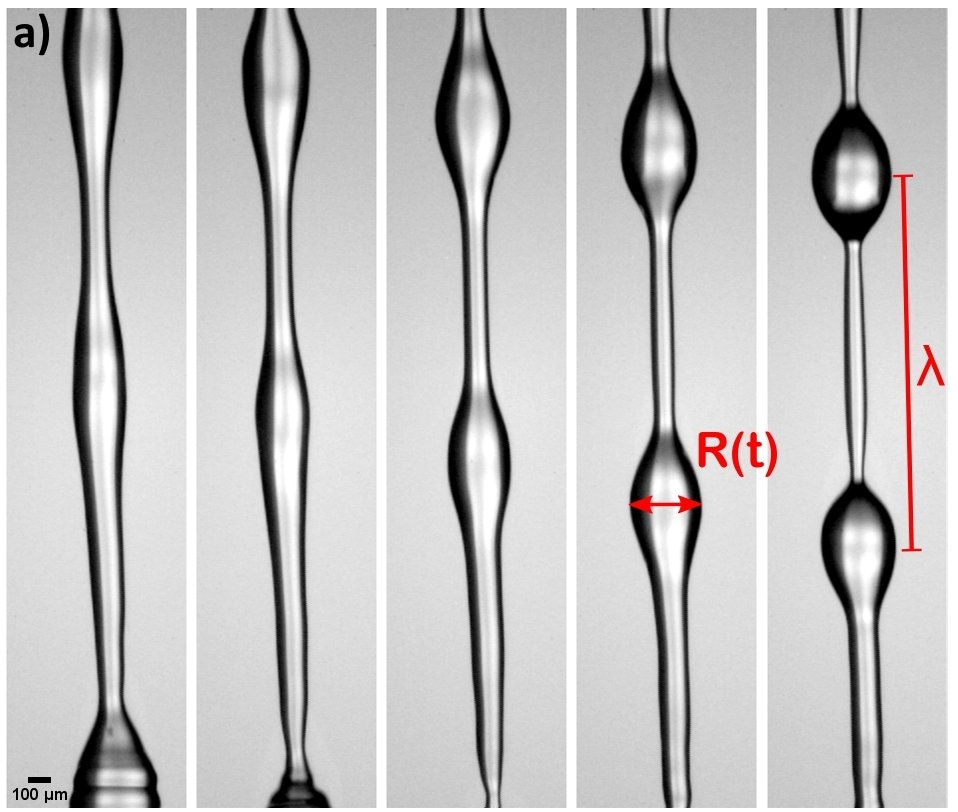}}
  \subfigure{\includegraphics[width = 0.48\textwidth]{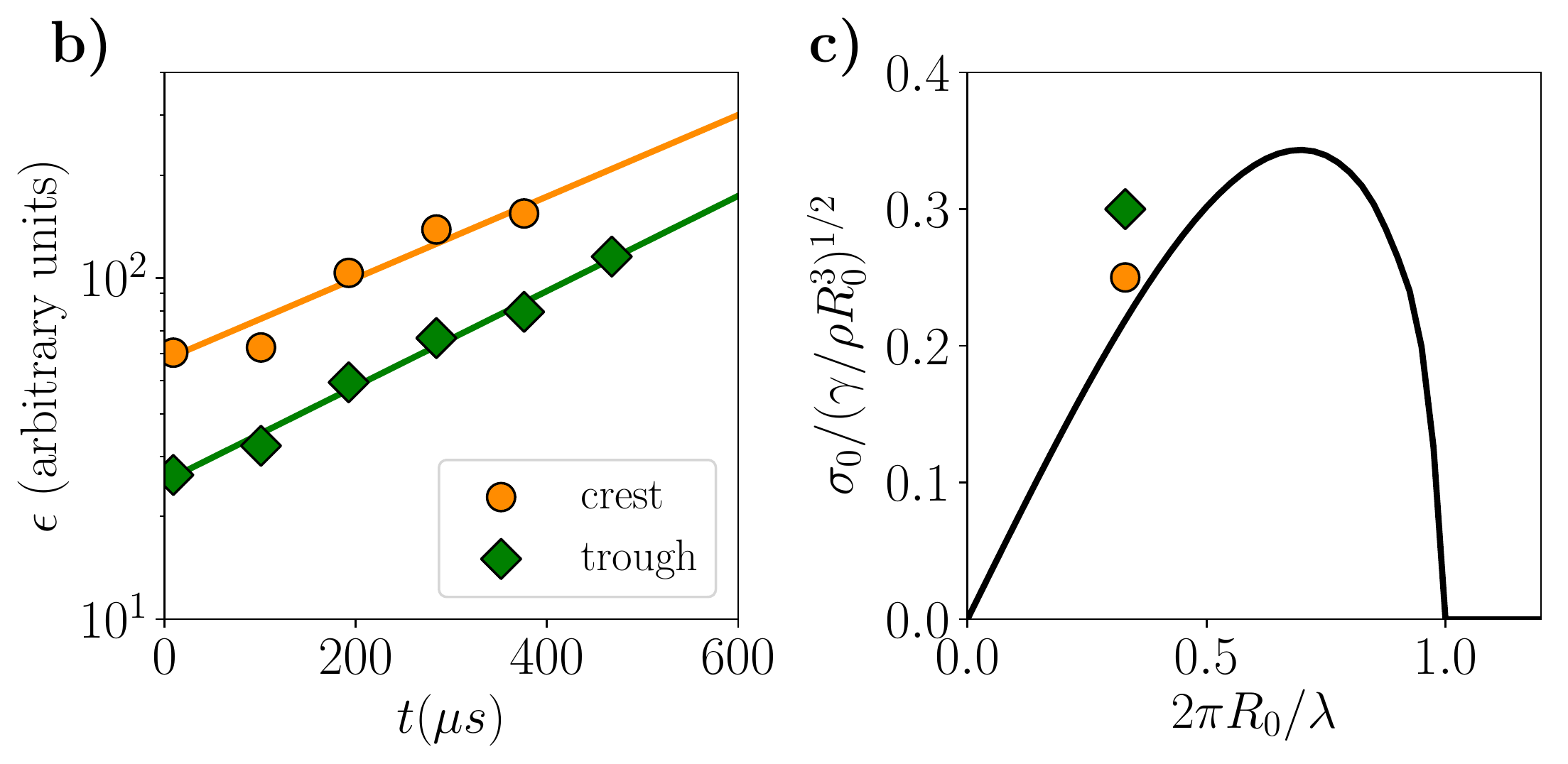}}
\caption{a) Typical image sequence recorded with a microscopic objective (10900 fps) of the breakup of a jet resulting from a droplet impact on a single-row metallic mesh. Here, $\lambda = 1.6 \;mm$. 
b) Swelling (of the crest) and thinning (of the trough) as a function of time determined from $R(t)= R_{0} \pm \epsilon(t)$, where $R_{0}$ is estimated to be $170\; \mu m$. An exponential fit of $\epsilon(t)$ gives a growth rate $\sigma_{0} = 2.8 \;m/s$ for the crest and $\sigma_{0} = 3.2 \;m/s$ for the trough. 
c) Comparison of the measured growth rates $\sigma_{0}$ and wavelength $\lambda$ with the dispersion relation as taken from \citep{chandrasekhar1970hydrodynamic}. The growth rates are nondimensionalized with the characteristic growth rate $\sqrt{\gamma/\rho R_{0}^{3}}$, where $\gamma$ is the surface tension and $\rho$ the density. Although the errors in this type of measurements are typically large, still all points lie on the left side of the maximum. One major issue is that $R_{0}$ is not as well-defined as for normal capillary jets due to the stretching of the ligaments.     \footnotesize }  
\label{fig:growing_perturbation}
\end{figure}

These observations show much resemblance with experiments on capillary jets with imposed perturbations \citep{rayleigh1879capillary,crane1964effect,donnelly1966experiments, rutland1971non}. The instability of capillary jets has been extensively investigated \citep{rayleigh1879capillary,eggers2008physics,crane1964effect,lin1998drop,donnelly1966experiments,rutland1971non,rutland1970theoretical,eggers1997nonlinear,goldin1969breakup,mansour1990satellite}. For Newtonian fluids, the breakup of a capillary jet is the result of the exponential growth of initial perturbations, where the growth rate depends on the perturbation wavelength as given by the dispersion relation in \citep{donnelly1966experiments}. These jets are very sensitive: even when much care is taken to remove any perturbations, ambient noise sources such as small vibrations and sound waves will determine the breakup of the jet \citep{lafrance1977capillary}. In our experiments ambient noise is not the source of the observed disturbances, since the peaks of the waves are always at the same location when the experiment is repeated. Moreover, the long wave disturbances only appear for impact velocities of $v \gtrapprox 2 \;m/s$. This suggests that the droplet impact itself causes vibrations that eventually lead to the final breakup pattern. If these vibrations would be broadband, the fastest growing wavelength, which has wavenumber $x =2\pi R_{0}/\lambda= 0.697$ with $R_{0}$ the jet radius, would be the one observed. However, in our experiments the wavelengths are significantly larger than that, leaving us to conclude that these vibrations have a limited spectral range. This also implies that the breakup is sensitive to changes of the set-up and might be hard to reproduce. Indeed, when spanning the same $150\; \mu m$ fabric with a different tension over the two pillars, we find that the wavelength shown in Fig.~\ref{fig:breakup_sequence} can change as much as a factor of two. By taking high-speed (10900 fps) microscopic images of the breakup of the jets for a metallic single-row mesh, we were able to measure the growth rate of the instabilities (Fig.~\ref{fig:growing_perturbation}).
By determining the change of the radii ($R(t)$) of both the crests and the troughs, compared with the initial radius $R_{0}$, the growth of the perturbation, $\epsilon(t) =|R(t)-R_{0}| = \epsilon_{0} e^{\sigma_{0} t}$, could be measured. From the exponential fit of $\epsilon(t)$ (Fig.~\ref{fig:growing_perturbation}b) we find a growth rate of $\sigma_{0} = 2.8 \;ms^{-1}$ and $\sigma_{0} = 3.2 \;ms^{-1}$ for the crest and trough, respectively. We find that the growth rates roughly agree with the dispersion relation for capillary jets (see Figure \ref{fig:growing_perturbation}c). It should be noted however that, unlike disturbed capillary jets, the initial radius of the jet, $R_{0}$, in our experiments is not well-defined. Due to stretching, the ligament diameter changes strongly over time, and also varies considerably over the length of the ligament. This together with other experimental uncertainties induces large errors. Still, for all jets, the measurement points lie consistently on the left hand side of the maximum growth rate depicted in Fig.~\ref{fig:growing_perturbation}c.
\newpage

\subsection{Drop size distribution}
\begin{figure}[!h]
\centering
\includegraphics[width = 0.48\textwidth]{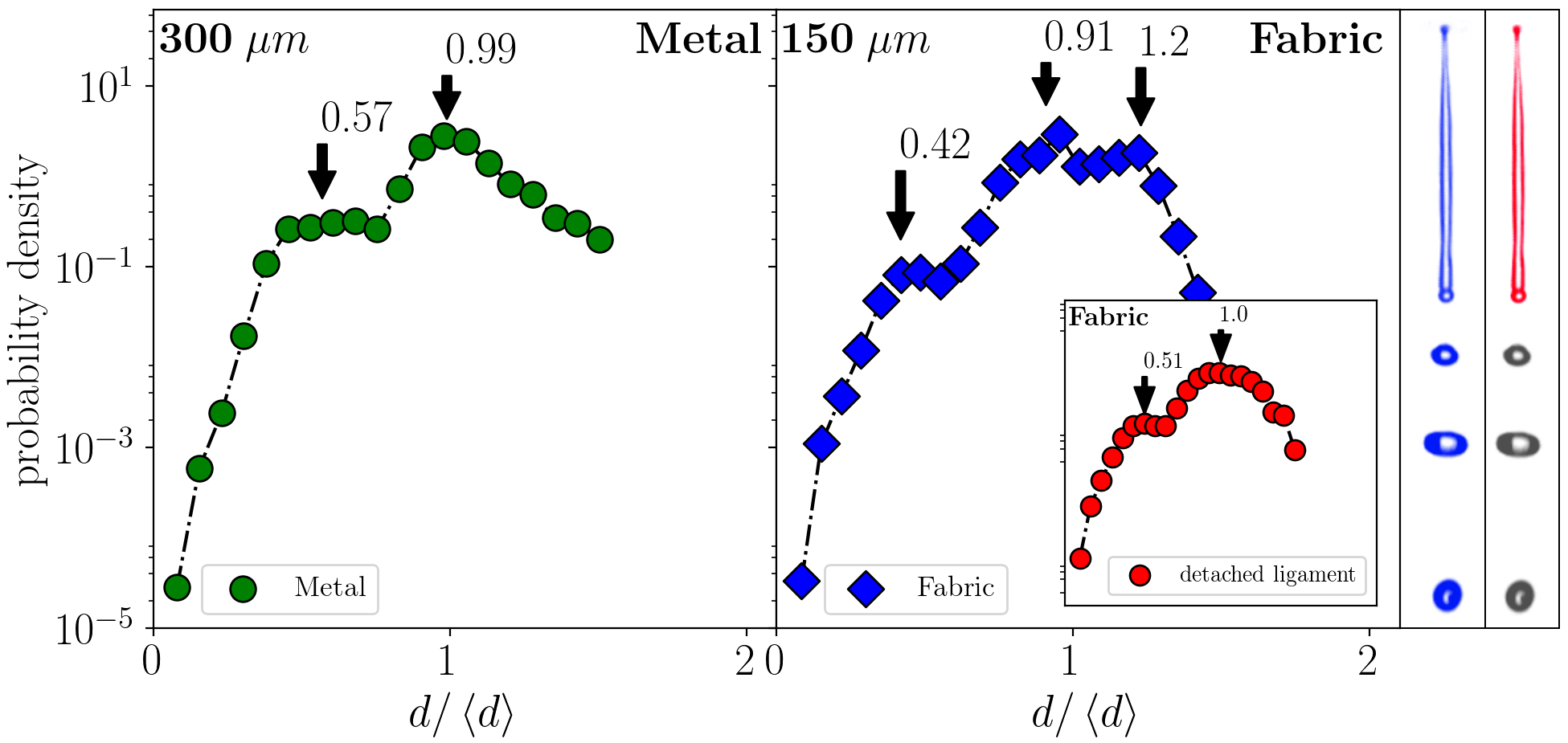}
\caption{Rescaled drop size distributions for a single-row metallic mesh (left) and a single-row polyester fabric mesh (right). The inset plot (right) shows the distribution of only droplets coming from detached ligaments. The difference between the blue and red distributions is illustrated by the blue and red snapshots on the right: by limiting the analysis to droplets created by ligaments breaking up, the lower three droplets seen in the snapshots get excluded, and the third peak in the droplet size distribution disappears. Clearly this peak in the original (red) distribution originated from the jetting part of the spray formation.   \footnotesize }
\label{fig:distribution_fabric_metal}
\end{figure}

Figure \ref{fig:distribution_fabric_metal} shows the rescaled droplet size distribution for the 300 $\mu m$ single-row metallic mesh and the $150\;\mu m$ single-row polyester fabric. In both cases there is a main peak and a smaller satellite peak. In the drop size distribution for the fabric there is also a small third peak visible at $d/\left<d\right> = 1.2$. When restricting the distribution to droplets coming from detached ligaments, the third peak disappears (see inset of Fig. \ref{fig:distribution_fabric_metal}). Closer examination reveals that this peak originates from $\sim 3$ droplets created during the pure jetting stage of the droplet impact, when the ligament did not go through a thinning process. Fig.~\ref{fig:overview_impacts}b confirms that the second wave of droplets are indeed significantly bigger. 

The first peak corresponds to the satellite droplet formation that originates from thin ligaments that connect the crests of the long wave disturbances during destabilization as previously described. These long wave disturbances are due to the droplet impact itself and therefore cause the droplet size distribution to be much broader. The second peak are the main droplets coming from the crest of the disturbances. 

The distributions created by single-row meshes seem to be quite different from those observed for full meshes (Fig.~\ref{fig:distribution_whole_mesh_single_velocity}). The main difference is that for the full mesh case there is no distinct satellite peak visible. Instead there is a smooth excess of small droplets compared to predictions for non-corrugated ligaments. This can be explained by the fact that for the full meshes, many ligaments of different sizes and different injection histories are created, because of which the satellite peak is spread out. Secondly, for the full meshes, the droplets are measured further down the impact zone (necessary to reduce overlap between droplets), giving droplets sufficient time to coalesce (see following section); this recombination also reduces the satellite peak considerably. 

While predicting the droplet sizes for a single-row mesh is already difficult due to the complicated and sensitive jet dynamics, for the full mesh there are several additional factors that influence the size distribution, such as the many differences in ligament size and injection speeds, possible merging of jets, and the coalescence of droplets after fragmentation. It is however clear that because of satellite drop formation, considerably more small droplets are created than one would expect from the pore size.

\subsection{Coalescence}
\label{sec:coalescence}
To understand what determines the shape of the drop size distributions, one needs to know what controls the size and breakup of the ligaments created by the droplet impact. However, after this short ligament fragmentation period, there is a secondary process that changes the size distribution significantly. Due to relative velocities between droplets after fragmentation, droplets frequently coalesce after separation. This phenomenon is intrinsic to the system, since droplets that originate from the same ligament travel along the same line, thereby facilitating coalescence. In other experiments such as the formation of stretched ligaments by the withdrawal of a tube from a liquid surface, coalescence could also take place, but has not been reported. This is probably also due to the fact that droplets will fall back on the free surface before a significant amount of coalescence events could have taken place. For disturbed jets however, this is a known phenomenon \citep{eggers1997nonlinear}. 

\begin{figure}[!htb]
\centering
\includegraphics[width = 0.48\textwidth]{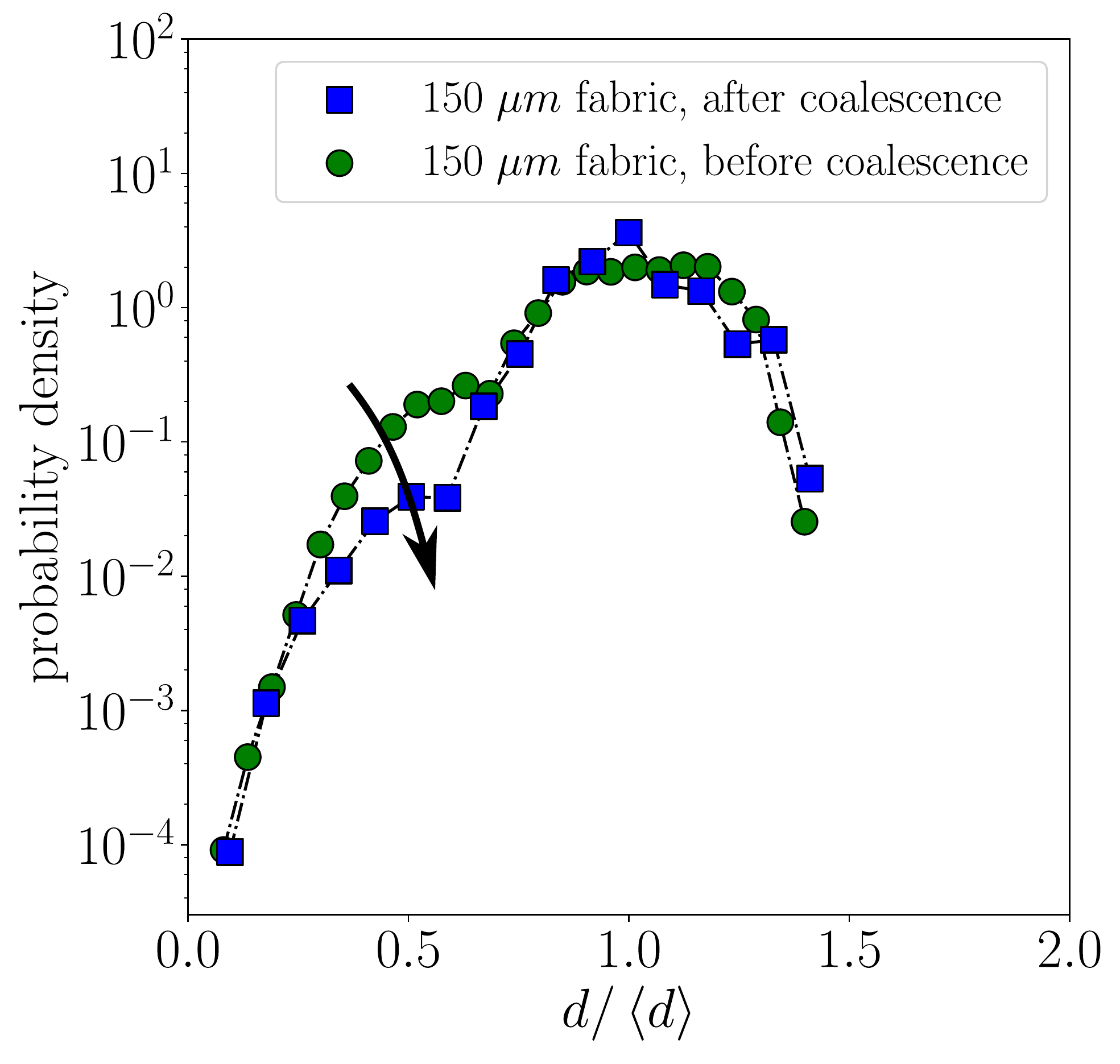}
\caption{Droplet size distributions of a single-row $150\;\mu m$ mesh (polyester fabric) measured directly after fragmentation, with no coalescence occurring, and after a waiting period so coalescence could take place. The relative amount of small droplets is decreased due to coalescence as indicated by the arrow.  \footnotesize }
\label{fig:dis_before_after_coalescence}
\end{figure}

From the high-speed footage, e.g. Fig.~\ref{fig:breakup_sequence_metallic_mesh}, we observe that due to coalescence the amount of small droplets is strongly reduced. The droplet size distribution is therefore different if it is measured further away from the impact zone, as can been seen in Fig.~\ref{fig:dis_before_after_coalescence} which shows droplet size distributions for single-row meshes. The first peak associated with satellite-like droplets is clearly reduced. Since for most applications droplet sizes would be measured further away from the impact zone, one can expect coalescence to play an important role in determining the observed size distribution.

\section{Simulations}
\label{sec:simulations}

\begin{figure*}[htbp!]
\centering
\includegraphics[width = 1.0\textwidth]{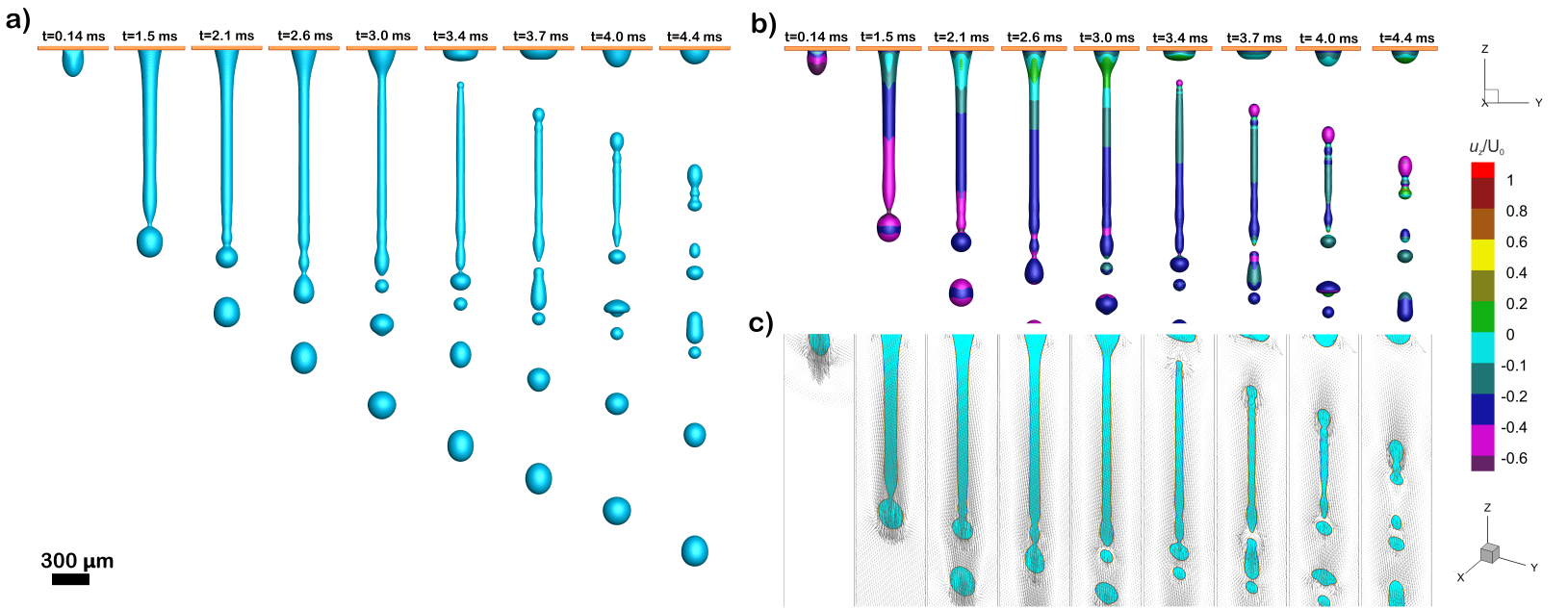}
\caption{a) Sequential images of the simulation of water injected through a 300 $\mu m$ hole at an exponentially slowing injection speed. b) Sequential images as in a) where the color indicates $u_{z}/U_{0}$, with $u_{z}$ the velocity component in the $z$-direction and $U_{0}$ is the injection speed at $t=0\; ms$. Even though there is much stretching before detachment, after detachment the velocities are relatively equal over the length of the ligament. This agrees with experimental results, where sequential images of the detached ligaments show very little contraction (see Figs~\ref{fig:breakup_sequence_metallic_mesh} and \ref{fig:breakup_sequence}). c) Velocity vectors of the middle plane of the computational domain for both liquid and air (vapor). There is a layer of air moving with the jet, that at first has a gradient in the $z$-direction, but at detachment becomes homogeneous.    \footnotesize }
\label{fig:simulations_ligament_formation}
\end{figure*}

To provide insight into whether perturbations indeed originate from the drop impact itself, and how the breakup mechanism would look like if such perturbations were absent (something that is not experimentally attainable), we perform numerical simulations using the recently proposed entropic lattice Boltzmann method for two-phase flows \citep{PhysRevLett.114.174502,mazloomi2015entropic}. To this end, the fragmentation process of a liquid jet through a single pore is modeled by injecting liquid with an exponentially slowing injection speed through a hole having the same size as the pore. A liquid flux boundary condition is implemented on the top surface of the single pore, allowing the liquid to be pushed through the pore, but also taking into account the effect of the yarn diameter (the details in implementation of the flux boundary condition can be found in \citep{mazloomi2018dynamics}). The injection speed used in simulations is assumed to be equal to the experimentally determined speed with which the top of the droplet moves downward (Fig.~\ref{fig:exp_injection_speed}). The liquid properties such as density and surface tension used in simulations are the same as reported in \citep{moqaddam2017drops}. The liquid viscosity $\mu$ is set according to the Ohnesorge number ($Oh = \mu /\sqrt{\rho \gamma D_{0}}$) for the water droplet used in experiments. The pore size and the yarn diameter are also set to match the experiment by keeping the aspect ratio of the droplet diameter to the pore size/yarn diameter the same as in the experiments. We also consider a solid-liquid contact angle of $\sim 70^\circ$ comparable to that of our experiments. Since the droplet size used in experiment is smaller than the capillary length for a water droplet, one can neglect the effect of gravity in the simulations. The simulation results are reported after studying the grid independence.

Figure \ref{fig:simulations_ligament_formation}a shows sequential images of the fragmentation process for a liquid jet through a 300 $\mu m$ hole obtained from numerical simulations (See also Figs S4, S5, and S6 in the Supplemental Materials). The observed sequence is found to be in good agreement with those seen in our experiments of impacting drops on single-row meshes. The imposed liquid flux boundary condition pushes liquid through the pore, resulting in the formation of a liquid jet which is followed by the breakup at the tip with a single droplet having a size of the order of the pore size ($t=2.1 \; ms$). By slowing down the injection speed, due to internal flow inside the liquid jet, the resultant ligament stretches and becomes thinner until the injection speed is so slow that the ligament detaches from the pore ($t=3.4\; ms$); finally, the detached narrow liquid ligament destabilizes and breaks up into smaller droplets ($t=4.4 \;ms$). Simulations allow us to visualize the quantities that are more difficult to be observed by experiments. Figs \ref{fig:simulations_ligament_formation}b and c show the velocity contour ($u_{z}/U_{0}$, where $u_{z}$ is the velocity value in $z$-direction and $U_{0}$ is the injection speed magnitude at $t=0 \;ms$) and the velocity vectors for the middle plane of the liquid jet, respectively. It can be seen that at the liquid neck that connects about-to-form drops with the rest of the ligament, the velocity value is relatively large, leading to liquid pinch-off and droplet formation. Although after ligament detachment ($t\geq 3.4 \;ms$) a larger downwards velocity at the ligament tail is observed, the ligament experiences little contraction as the downward velocity of the rest of the ligament is still relatively large. Visualization of the velocity vectors also shows that the liquid jet carries a relatively thick layer of air as injection proceeds. Furthermore, the velocity field within the ligament and the fragmented droplets obeys mostly the direction of the initial injection speed. Simulations show a rapid reversal flow or circulation at the ligament tip right after the liquid pinch-off occurs (See Movie S2 in the Supplemental Materials). Simulations also exhibit coalescence between small fragmented droplets due to their small relative velocities, similar as observed in our experiments.

\section{Discussion and conclusions}
We found that the fragmentation of a droplet impacting a single-row mesh is controlled by a jetting instability, where initial perturbations determine the final breakup of the jets in a deterministic fashion. The source of the perturbations is the droplet impact itself, causing regular long-wavelength disturbances on the jet's surface that exponentially grow to form thick blobs at the crests of these waves. These blobs are connected by thin ligaments that break up to form satellite droplets, leading to a bimodal size distribution. Due to relative velocities between droplets after fragmentation, a secondary process of coalescence significantly reduces the amount of smaller droplets. 

The droplets coming from the impact on a full mesh have a similar distribution as droplets coming from just a single row of pores. Both have an excess of small droplets, however a distinct satellite peak is missing. A droplet impacting on a full mesh creates many ligaments of different sizes with different injection histories. Together with a secondary process of coalescence, this causes the distribution of smaller droplets to be more spread out. 

We investigated several factors that affect the fragmentation of the impacting droplet, such as impact velocity, wetting properties, and mesh rigidity. Even though most parameters affect the formation and breakup of the created jets, usually the same characteristic satellite drop formation is observed. Therefore, the most important factor in the fragmentation seems to be the perturbations during the injection process. This factor is however also the most difficult to control. It could well be that with a change of set-up, other frequencies will be observed, thereby changing the size distribution. It remains somewhat puzzling for example, why with this set-up only slow-growing modes are excited, when no specific effort was made to reduce noise sources.   

Simulations show that the injection process can be viewed as a simple system with cylinder and piston, where the piston height decreases exponentially with time, pushing liquid through a hole on the bottom of the cylinder. The sprays created in this manner look very similar to those observed experimentally, with the important difference that the detached ligament is free of disturbances. If perturbations would be added, we expect to recover the basic fragmentation mechanism of a droplet impacting a mesh.  

Droplet fragmentation due to impact with a mesh seems a simple way of reducing the droplet size, since the droplet size is controlled by the dimensions of the pores. However, when the drop size distribution is properly rescaled, this spray formation process performs rather poor compared to other atomization methods. Satellite drop formation is the main reason for a broad size distribution, something that is not uncommon in the destabilization of capillary jets such as seen in this process. Still, many important system properties have yet to be explored such as viscosity, surface tension, and pore shape. Moreover, a more extensive investigation into the nature of the perturbations could point to ways to improving the spraying properties of this particular technique.

\section*{Acknowledgements}
This work is part of the Industrial Partnership Program Hybrid Soft Materials that is carried out under an agreement between Unilever Research and Development B.V. and the Netherlands Organisation for Scientific Research (NWO).

A.M.M., D.D., and J.C. acknowledge the support by the Swiss National Science Foundation (Project no. 200021\_175793). The computational resources were provided by the Swiss National Supercomputing Center (CSCS) under project number s823. The authors thank Emmanuel Villermaux for fruitful discussions.

\newpage{}
%

\end{document}